%% file: main.tex
\documentclass{article}

\usepackage{dilab_arxiv}

\usepackage{amsfonts}
\usepackage{ifthen}
\usepackage{textcomp}
\usepackage{mathtools}
\usepackage{bbm,dsfont} 
\usepackage{algorithm}
\usepackage{wrapfig}
\usepackage{algorithmic}
\usepackage{multirow} 
\usepackage{tikz}
\usetikzlibrary{arrows.meta,positioning}

\usepackage{CJKutf8} 

\usepackage[capitalise,nameinlink]{cleveref}

\let\standardEqref\eqref

\input{math_commands}

\let\eqref\standardEqref

\theoremstyle{plain}

\theoremstyle{definition}

\theoremstyle{remark}

\newcommand{\compilehidecomments}{false}
\ifthenelse{ \equal{\compilehidecomments}{true} }{%
	\newcommand{\yu}[1]{}
    \newcommand{\longbo}[1]{}
    
    \colorlet{ruishuocolor}{black}
}{
	\newcommand{\yu}[1]{{\color{cyan}[\text{Yu:} #1]}}
    \newcommand{\longbo}[1]{{\color{orange}[\text{Longbo:} #1]}}
    \definecolor{ruishuocolor}{RGB}{0,155,80}
    
}

\title{Abstraction Agent}
\runningtitle{Abstraction Agent}
\date{\today}

\paperlogo{\includegraphics[height=1.5cm]{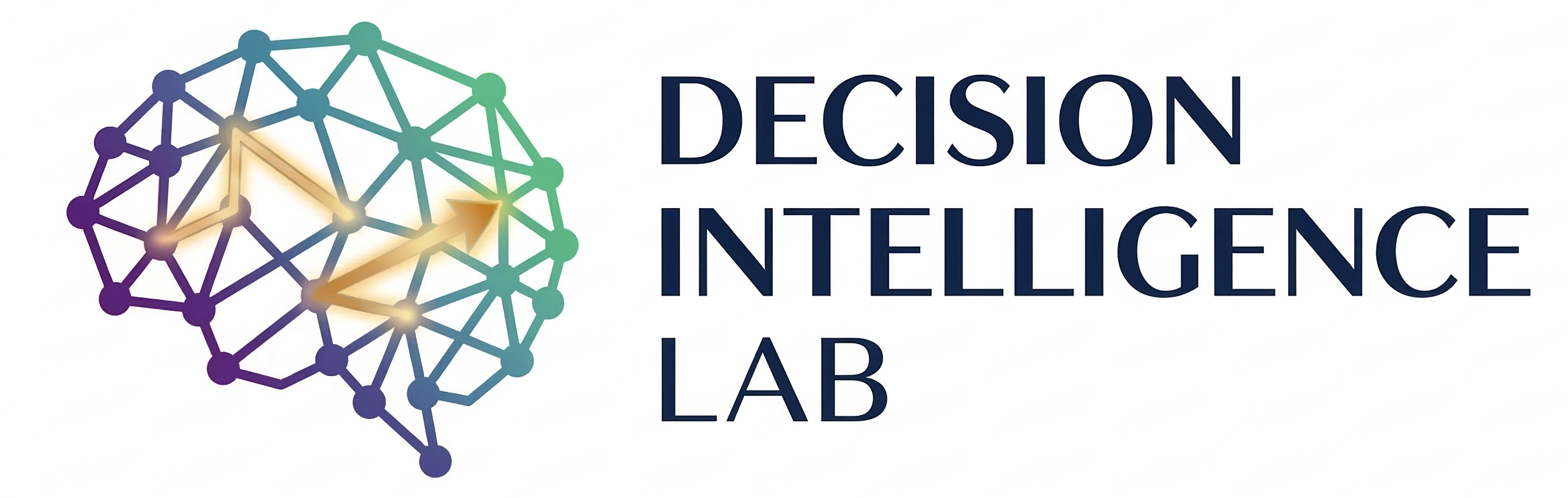}}

\author{
  Boning Li$^{1}$ and Longbo Huang$^{1\,\text{\faEnvelope}}$
  \\[0.3em]\normalfont
  $^1$Institute for Interdisciplinary Information Sciences, Tsinghua University
  \\
  \text{\faEnvelope}\ Correspondence: longbohuang@tsinghua.edu.cn 
}

\begin{document}

\maketitle
\thispagestyle{fancy}

\begin{abstract}
Information abstraction, which groups strategically similar private states into a tractable number of buckets, is essential for scaling game-solving algorithms to large imperfect-information games. Constructing effective abstractions, however, has traditionally required domain-specific evaluators such as hand-strength calculators or equity estimators, which demand expert knowledge and engineering effort and are unavailable for most less-studied games. We propose the \emph{Abstraction Agent}, a zero-shot pipeline that uses a large language model (LLM) to discover continuous strategic features from a natural-language game description, score private states on these features, and cluster them into abstraction buckets, without any game-specific evaluator, training data, or game-tree traversal during abstraction construction. The pipeline runs in four phases: feature discovery with calibration anchors, batched private-state scoring, correlation-based feature selection, and $k$-means clustering. The resulting abstractions reduce lifted-strategy exploitability by up to 62\% relative to an expected-hand-strength baseline on heads-up no-limit Texas hold'em (HUNL) turn endgames, and beat a scalar rank baseline at every granularity on ROVER Trials, an original game absent from any pretraining corpus. Beyond these quantitative benchmarks, the pipeline transfers with unchanged prompts to four-card Pot-Limit Omaha, HUNL preflop and flop, and Riichi Mahjong, where the discovered features track each game's recognized strategic concepts. This is \emph{structured knowledge elicitation}: converting implicit strategic knowledge in LLM parameters into explicit numerical features for downstream algorithmic computation. The code is available at \url{https://github.com/lbn187/AbstractionAgent}.
\end{abstract}

\input{sections/introduction}
\input{sections/related_work}
\input{sections/preliminaries}
\input{sections/method}
\input{sections/experiments}
\input{sections/results}
\input{sections/analysis}
\input{sections/conclusion}

\bibliography{references}
\bibliographystyle{dilab_ref}
\makeappendixtoc

\input{sections/appendix}

\end{document}

%% file: math_commands.tex
\usepackage{amsmath,amsfonts,bm,bbm}

\Crefname{ALC@unique}{Line}{Lines}

\Crefname{algorithm}{Algorithm}{Algorithms}
\Crefname{assumption}{Assumption}{Assumptions}
\Crefname{lemma}{Lemma}{Lemmas}
\Crefname{proposition}{Proposition}{Propositions}
\Crefname{corollary}{Corollary}{Corollaries}
\Crefname{theorem}{Theorem}{Theorems}
\Crefname{definition}{Definition}{Definitions}
\Crefname{remark}{Remark}{Remarks}

\Crefformat{equation}{Eq. #2(#1)#3}
\Crefrangeformat{equation}{Eqs. #3(#1)#4 to #5(#2)#6}
\Crefmultiformat{equation}{Eqs. #2(#1)#3}{ and #2(#1)#3}{, #2(#1)#3}{ and #2(#1)#3}
\Crefrangemultiformat{equation}{Eqs. #3(#1)#4 to #5(#2)#6}{ and #3(#1)#4 to #5(#2)#6}{, #3(#1)#4 to #5(#2)#6}{ and #3(#1)#4 to #5(#2)#6}

\def\eqref#1{equation~\ref{#1}}

\def\1{\mathbbm{1}}

\DeclareMathAlphabet{\mathsfit}{\encodingdefault}{\sfdefault}{m}{sl}
\SetMathAlphabet{\mathsfit}{bold}{\encodingdefault}{\sfdefault}{bx}{n}

\newif\ifsup\supfalse
\suptrue

%% file: sections/introduction.tex
\section{Introduction}

Large language models (LLMs) have shown strong capabilities in reasoning and planning across diverse tasks \citep{wei2022chain, kojima2022large}. Recent work explores LLMs as game-playing agents \citep{meta2022diplomacy, gandhi2024strategic, DBLP:journals/corr/abs-2309-17277}, yet in complex strategic games their play remains limited by imprecise probability estimation and inconsistent decision-making \citep{huang2024pokergpt, zhuang2024pokerbench, lin2026how}. We therefore propose using LLMs not as players, but as \emph{feature engineers} that identify the strategic dimensions along which game states should be compared. This division of labor plays to the LLM's strength of articulating qualitative strategic structure in natural language, while delegating the quantitative equilibrium computation to an exact game solver, where LLMs are unreliable.

In imperfect-information games, where players must act under uncertainty about opponents' private states, \emph{information abstraction} is essential for computational tractability. Abstraction groups strategically similar private states into buckets, enabling solvers like counterfactual regret minimization (CFR) to operate on manageable game sizes \citep{zinkevich2007regret}. The quality of this grouping is a major determinant of solution quality: merging states with different strategic roles introduces exploitability.

Constructing effective abstractions has traditionally required domain-specific evaluators. In poker, hands are clustered by equity or expected hand strength (EHS), metrics that require hand evaluators, Monte Carlo rollouts, or expert-designed features \citep{johanson2013measuring}. Recent advances such as EVPA \citep{li2025evpa} and WEVA \citep{li2026weva} reduce the dependency on neural training or handcrafted features, but both still gather their abstraction signal by traversing the game tree. This traversal is computationally expensive for large games and, more fundamentally, presupposes that a game-specific solver or evaluator has already been implemented, which is often not the case for less-studied games. More broadly, many existing methods rely on at least one of: (1) game-specific evaluators, (2) extensive training data from game-tree traversal, or (3) closed-source implementations that make reproduction difficult \citep{brown2018superhuman, brown2019superhuman}.

This motivates our central question: \emph{can an LLM, given only a natural-language description of the game rules, construct useful information abstractions without game-tree traversal during abstraction construction?}

We provide initial evidence that this is possible by proposing an \emph{Abstraction Agent}, a zero-shot pipeline that uses LLM reasoning to discover and apply strategic features for information abstraction. By ``zero-shot'' we mean that the pipeline uses no game-specific training data, rollout evaluator, or game-tree traversal; it receives only a natural-language rules description and textual renderings of private states. The agent operates in four phases: (1) \textbf{Feature Discovery}, the LLM proposes continuous strategic features with calibration anchors from a game description; (2) \textbf{Private-State Scoring}, each holding is scored on these features via batched LLM calls; (3) \textbf{Feature Selection}, redundant features are filtered by correlation analysis; (4) \textbf{Clustering}, the resulting feature vectors are partitioned with $k$-means. The whole pipeline avoids domain-specific evaluators and manually designed scoring functions.

The key insight is that LLMs trained on broad text corpora encode useful strategic regularities about well-studied games. Concepts like ``nut potential,'' ``blocker value,'' and ``draw vulnerability'' exist in the LLM's knowledge as natural-language descriptions of strategic dimensions. Our pipeline elicits this knowledge as \emph{structured, continuous features} suitable for downstream computation. We call this process \emph{structured knowledge elicitation}: prompting an LLM to convert implicit domain knowledge into explicit, machine-readable numerical variables that can be consumed by non-neural downstream algorithms.

We evaluate on the two publicly available HUNL turn endgames from the Libratus release \citep{brown2018superhuman}, where the agent's abstraction reduces exploitability relative to EHS by up to 62\% at the finest granularity with the advantage growing monotonically in the bucket count. Because poker is heavily represented in pretraining corpora, we also evaluate on ROVER Trials, an original two-player zero-sum game absent from any pretraining corpus: given only the rules, the same pipeline beats the scalar rank baseline at every granularity and a 2D potential-aware baseline at most granularities, indicating that the agent \emph{constructs} features rather than retrieving memorized poker knowledge. The unchanged pipeline also produces strategically coherent abstractions in PLO4, HUNL preflop/flop, and Riichi Mahjong, games where evaluator-based abstraction signals are too expensive to compute or unavailable, at a cost linear in the number of holdings.

Our contributions are:
\begin{enumerate}
    \item \textbf{A zero-shot abstraction pipeline.} We propose the first LLM-based pipeline that converts a natural-language game description into a multi-dimensional information abstraction without hand evaluators, rollout simulators, or game-tree traversal during abstraction construction.
    \item \textbf{Quantitative gains over standard baselines.} On the two public HUNL turn subgames, the agent's abstraction reduces exploitability relative to EHS by up to 62\% at $K{=}100$, beating EHS in all $K{=}100$ runs without any domain-specific evaluator, with the advantage growing monotonically in the bucket count $K$; the gain requires a frontier backbone.
    \item \textbf{Construction beyond memorization, with cross-game portability.} On ROVER Trials, an original game absent from any pretraining corpus, the rules-only pipeline beats the scalar rank baseline at every granularity, showing genuine construction rather than retrieval of memorized poker features, and produces strategically coherent abstractions in PLO4, HUNL preflop/flop, and Riichi Mahjong without modification.
\end{enumerate}

%% file: sections/related_work.tex
\section{Related Work}

\paragraph{Information Abstraction in Games.}
Abstraction is a core technique for scaling imperfect-information game solving. In extensive-form games, information sets group states indistinguishable to a player; abstraction further merges information sets deemed strategically similar \citep{sandholm2015abstraction}. Early work on poker abstraction used hand rank bucketing \citep{shi2000abstraction, billings2003approximating}. Modern approaches cluster private hands by expected hand strength (EHS) or equity against a uniform opponent range \citep{johanson2013measuring, johanson2013evaluating}. Potential-aware abstractions \citep{gilpin2007potential, ganzfried2014potential} additionally consider how hand values change with future public cards. RL-CFR \citep{li2024rlcfr} uses reinforcement learning to optimize \emph{action} abstraction, complementing information abstraction. EVPA \citep{li2025evpa} trains neural networks on CFR traversal data to estimate per-hand features for online abstraction. WEVA \citep{li2026weva} avoids neural training but still requires game-tree traversal. Our work removes the requirement for game-specific evaluators and game-tree traversal entirely, using LLM-derived features instead.

\paragraph{CFR and Exploitability.}
Counterfactual regret minimization (CFR) \citep{zinkevich2007regret} and its sampling-based variants \citep{lanctot2009monte, li2026correlated} are standard algorithms for approximating Nash equilibria in two-player zero-sum games. Equilibria of such games can also be computed by linear programming, by first-order saddle-point methods such as the excessive gap technique \citep{kroer2018solving}, and by no-regret learning dynamics such as fictitious play; we adopt CFR for its scalability and its ubiquity in modern poker solving. Superhuman poker AI systems \citep{moravvcik2017deepstack, brown2018superhuman, brown2019superhuman} combine CFR with abstraction and real-time subgame solving. Recent work on parallelizing CFR \citep{li2026parallel} enables real-time solving on desktop hardware. We use CFR as the downstream solver and evaluate abstractions by the exploitability of the lifted average strategy \citep{waugh2009practical, johanson2013evaluating} and, where exact best responses are intractable, certify agent performance with anytime-valid estimators \citep{li2026av}. Downstream of equilibrium computation, agents may also deviate from equilibrium to exploit opponents under certified exploitability budgets \citep{li2026agents}, or replace static chip-value heuristics with computed continuations in tournament settings \citep{li2026icm}; both directions consume the abstractions studied here.

\paragraph{Domain-Knowledge-Free Abstraction.}
Deep CFR \citep{brown2019deepcfr} replaces tabular regret with neural function approximation trained from self-play traversals, avoiding manual feature design but requiring large amounts of game-specific sampling. WEVA \citep{li2026weva} eliminates neural training but still requires running CFR iterations on the unabstracted game. Our approach uses pretrained LLM knowledge as the source of abstraction signal, requiring neither game-specific training nor game-tree traversal, making it applicable to games where no solver implementation exists.

\paragraph{LLMs for Strategic Reasoning.}
We distinguish \emph{strategic} game playing, in which a player reasons about equilibria and adapts to an opponent, from \emph{non-strategic} play that follows fixed heuristics or imitates demonstrations \citep{gandhi2024strategic}. Recent work has explored LLMs as game-playing agents \citep{meta2022diplomacy, DBLP:journals/corr/abs-2309-17277, wang2026can}, strategic reasoners \citep{gandhi2024strategic}, and value estimators inside search \citep{karten2025pokchamp}. In poker, LLMs have been evaluated as zero-shot players \citep{DBLP:journals/corr/abs-2308-12466, huang2024pokergpt, zhuang2024pokerbench, peng2025can} with mixed results, reflecting the difficulty of reliable equilibrium-aware play; systematic diagnoses attribute the failures to heuristic shortcuts and misreadings of the game state \citep{lin2026how}. Concurrent work on structured prompting \citep{li2026pokerskill, wang2026solver} demonstrates that carefully structured prompts can activate latent poker strategy in frontier LLMs. While these works use LLMs as \emph{players} or online evaluators, our work uses LLMs as \emph{feature engineers} for offline abstraction construction, a complementary role that feeds into traditional game-solving pipelines.

\paragraph{LLMs as Feature Engineers.} Explicit structural features are known to complement learned representations \citep{cao2015grarep,li2021distance}. The idea of using LLMs to propose features has emerged in tabular learning \citep{hollmann2023caafe} and AutoML contexts. More broadly, LLMs can translate natural-language rules into formal game structures: extensive-form game representations \citep{deng2025natural}, executable world models for general game playing \citep{lehrach2025code}, and dynamic deduction that prunes actions and states during equilibrium learning \citep{zhang2026generative}. Our work applies this paradigm to game-theoretic abstraction, where features must satisfy game-theoretic desiderata (spread, independence, coverage) that differ from standard ML feature quality metrics. The key technical challenge is eliciting \emph{calibrated numerical outputs} from LLMs; we address this through calibration anchors and constrained output formatting.

%% file: sections/preliminaries.tex
\section{Preliminaries}
\label{sec:prelim}

We briefly fix the standard game-theoretic objects used throughout the paper.

\paragraph{Extensive-form games.}
A two-player zero-sum \emph{extensive-form game} is a tuple $\mathcal{G} = \langle \mathcal{N}, \mathcal{H}, \mathcal{Z}, \mathcal{A}, P, \{\mathcal{I}_i\}, \{u_i\} \rangle$. $\mathcal{N}=\{1,2\}$ is the set of players (a chance player $c$ models stochastic events such as dealing cards). $\mathcal{H}$ is the set of \emph{histories} (sequences of actions from the root); $\mathcal{Z}\subseteq\mathcal{H}$ are \emph{terminal} histories. At a non-terminal history $h$, the player to act $P(h)\in\mathcal{N}\cup\{c\}$ chooses from the available actions $\mathcal{A}(h)$. Each terminal $z\in\mathcal{Z}$ yields utility $u_i(z)\in\mathbb{R}$ to player $i$; the game is \emph{zero-sum} when $u_1(z)=-u_2(z)$.

\paragraph{Information sets.}
Because players act under uncertainty about hidden information (e.g., the opponent's cards), the decision histories of player $i$ are partitioned into \emph{information sets} $\mathcal{I}_i$. Two histories in the same information set $I\in\mathcal{I}_i$ are indistinguishable to $i$, so the same actions $\mathcal{A}(I)$ are available and $i$ must behave identically at both. An information set is identified by the player's \emph{private type} (its private state, e.g., its hole cards) together with the observed public history.

\paragraph{Strategies and equilibrium.}
A \emph{behavioral strategy} $\sigma_i$ assigns to each $I\in\mathcal{I}_i$ a distribution over $\mathcal{A}(I)$. A \emph{strategy profile} is $\sigma=(\sigma_1,\sigma_2)$, and $u_i(\sigma)$ denotes player $i$'s expected utility under $\sigma$ (taking expectations over chance and both players' randomization). A \emph{best response} for $i$ to the opponent profile $\sigma_{-i}$ attains $\max_{\sigma_i'} u_i(\sigma_i',\sigma_{-i})$. A \emph{Nash equilibrium} is a profile where neither player can gain by deviating.

\paragraph{Exploitability.}
The quality of a profile $\sigma$ in a two-player zero-sum game is measured by its \emph{exploitability}, the average amount an optimal adversary gains against each player:
\begin{equation}
\label{eq:expl}
\mathrm{expl}(\sigma) = \tfrac{1}{2}\sum_{i\in\mathcal{N}} \Big( \max_{\sigma_i'} u_i(\sigma_i',\sigma_{-i}) - u_i(\sigma) \Big).
\end{equation}
A Nash equilibrium has $\mathrm{expl}(\sigma)=0$; larger values indicate strategies further from optimal. We report exploitability as a fraction of the pot.

\paragraph{Information abstraction.}
Solving a large game exactly is intractable, so an \emph{information abstraction} $\phi$ groups the private types reaching a given public state into $K\ll N$ \emph{buckets}, $\phi:\{h_1,\dots,h_N\}\to\{1,\dots,K\}$. A solver computes a strategy $\sigma^\phi$ on the smaller abstract game, which is then \emph{lifted} back to the original game by forcing all private types in the same bucket to share the same action distribution. Because merged types must behave identically, a poor grouping raises the exploitability of the lifted strategy; the abstraction quality we evaluate is exactly $\mathrm{expl}$ of this lifted profile, computed by an exact best response over the unabstracted game.

%% file: sections/method.tex
\section{Method}

\begin{figure*}[t]
    \centering
    \includegraphics[width=\textwidth]{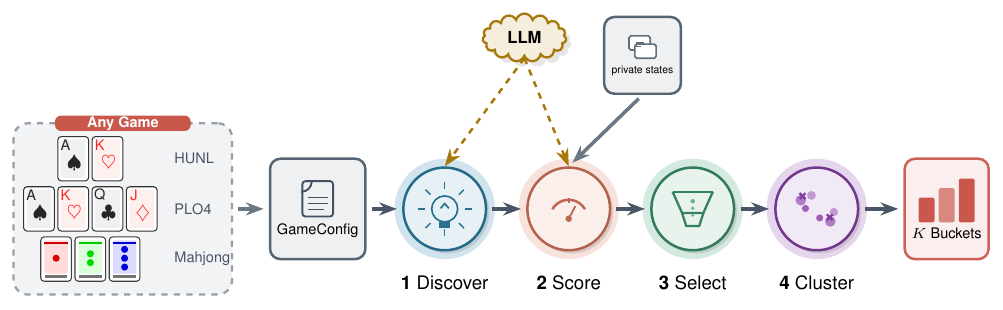}
    \caption{The Abstraction Agent pipeline. Given a textual \texttt{GameConfig} and a set of private holdings, an LLM discovers strategic features (Phase 1), scores each holding on them (Phase 2), redundant dimensions are filtered (Phase 3), and holdings are clustered into $K$ buckets (Phase 4). The pipeline uses no hand evaluators, training data, or game-tree traversal, and the same code applies to poker, Omaha, and Mahjong unchanged.}
    \label{fig:pipeline}
\end{figure*}

We propose an \emph{Abstraction Agent} that builds an information abstraction for an imperfect-information game through a four-phase pipeline (Figure~\ref{fig:pipeline}). The agent is game-agnostic: all domain knowledge enters through a declarative \texttt{GameConfig} that gives natural-language descriptions of the game rules and context. No hand-strength calculators, equity estimators, or rollout simulators are used.

\paragraph{Problem setting.}
Building on Section~\ref{sec:prelim}, we construct an information abstraction $\phi$ at a fixed public state $c$ (the community cards in poker, the visible discards in Mahjong). At this public state each player holds one of $N$ \emph{private types} $h_1,\dots,h_N$; we call a private type interchangeably a \emph{private state} or \emph{holding} (hole cards in poker, concealed tiles in Mahjong). The abstraction $\phi:\{h_1,\dots,h_N\}\to\{1,\dots,K\}$ assigns each private type to one of $K$ buckets; a solver (CFR) computes a strategy over the buckets, which is then \emph{lifted} back to the original game as in Section~\ref{sec:prelim}, so private types in the same bucket share an action distribution. The agent's goal is a $\phi$ whose lifted strategy has low exploitability (Eq.~\ref{eq:expl}) \emph{without any game-specific evaluator}.

\paragraph{Design rationale.}
An alternative is to have the LLM bucket holdings directly, but that requires every holding in one context for consistent grouping and a fresh generation for each $K$. Scoring against fixed features instead keeps each call local: the anchors pin each feature's endpoints and midpoint to concrete situations, so batches scored independently share a scale, and the global partitioning is left to $K$-means. The explicit features also remain readable before any solving, and Section~\ref{sec:analysis} traces abstraction quality back to individual features.

\paragraph{Phase 1: feature discovery.}
Given a \texttt{GameConfig} with the game rules and strategic context, the agent prompts the LLM to propose a set of continuous strategic features $\{f_1,\dots,f_M\}$ (typically $M\in[3,8]$). Each feature $f_j$ carries:
\begin{itemize}
    \item a \textbf{name} (e.g., \textsc{Nut\_Ceiling}, how close a holding is to the strongest possible hand on the current board);
    \item a \textbf{description} of what it measures;
    \item \textbf{calibration anchors} mapping scores 0.0, 0.5, and 1.0 to concrete game situations.
\end{itemize}
The prompt asks for features that are continuous in $[0,1]$, spread across the hand space (std $\geq 0.15$), mutually independent, and jointly covering the main strategic axes; when the game has future public cards it additionally requests future-card sensitivity. This phase needs a single LLM call, cached for reuse across $K$.

\paragraph{Phase 2: private-state scoring.}
For each private holding $h_i$ the agent prompts the LLM with the public context, the discovered feature definitions and anchors, and a batch of 20--30 holdings, and asks for a score vector $\mathbf{s}_i\in[0,1]^M$. The anchors provide a shared reference frame so scores stay consistent across batches. Holdings are processed in batches with retry logic and caching; a score that still fails to parse defaults to the neutral midpoint $0.5$, placing the holding in a central bucket rather than dropping it. In practice parse rates are at least 97.9\%, so the fallback rarely fires. This yields the feature matrix $\mathbf{S}\in\mathbb{R}^{N\times M}$.

\paragraph{Phase 3: feature selection.}
Not all discovered features are informative or independent. We drop columns with standard deviation below $\tau_\sigma=0.10$ (the LLM assigned near-constant scores), then, for each surviving pair with Pearson correlation $|r|>\tau_r=0.90$, drop the lower-variance member. The survivors are rescaled by dividing each column by its standard deviation (no mean-centering, since the anchors already provide a shared zero-point), giving $\mathbf{Z}\in\mathbb{R}^{N\times M'}$.

\paragraph{Phase 4: clustering.}
We partition the $N$ holdings into $K$ buckets with multi-restart $K$-means on $\mathbf{Z}$:
\begin{equation}
    \phi(h_i) = \arg\min_k \| \mathbf{z}_i - \boldsymbol{\mu}_k \|^2
\end{equation}
using $K$-means++ initialization, 10 restarts of 50 iterations each, keeping the partition with minimal within-cluster sum of squares. Algorithm~\ref{alg:agent} summarizes the pipeline.

\begin{algorithm}[t]
\caption{Abstraction Agent Pipeline}
\label{alg:agent}
\begin{algorithmic}[1]
\REQUIRE Game config $\mathcal{G}$, holdings $\{h_1,\ldots,h_N\}$, public state $c$, bucket count $K$
\ENSURE Abstraction $\phi: \{h_i\} \to \{1,\ldots,K\}$
\STATE \textbf{Phase 1: Feature Discovery}
\STATE $\{f_1,\ldots,f_M\} \leftarrow \text{LLM}(\mathcal{G}.\text{game\_description})$
\STATE \textbf{Phase 2: Private-State Scoring}
\FOR{each batch $B \subset \{h_1,\ldots,h_N\}$}
    \STATE $\mathbf{s}_i \leftarrow \text{LLM}(c, \{f_j\}, B)$ for $h_i \in B$
\ENDFOR
\STATE $\mathbf{S} \leftarrow [\mathbf{s}_1; \ldots; \mathbf{s}_N] \in \mathbb{R}^{N \times M}$
\STATE \textbf{Phase 3: Feature Selection}
\STATE Drop columns with $\text{std} < \tau_\sigma$
\STATE Drop correlated pairs with $|r| > \tau_r$
\STATE $\mathbf{Z} \leftarrow$ standardize surviving columns
\STATE \textbf{Phase 4: Clustering}
\STATE $\phi \leftarrow K\text{-means}(\mathbf{Z}, K)$ with multi-restart
\RETURN $\phi$
\end{algorithmic}
\end{algorithm}

%% file: sections/experiments.tex
\section{Experimental Setup}

\paragraph{HUNL Endgames.}
We evaluate on the two turn endgames publicly available from the Libratus subgame release \citep{brown2018superhuman}, which form the complete set of public turn subgames from that release rather than a hand-picked subset. Each subgame specifies a public board (4 community cards with one more to come), pot size, and private hands reaching the decision point. Turn boards are the most strategically complex street: hand rankings can change when the final card is revealed, creating multi-dimensional strategic value (current strength, draw potential, vulnerability) that a single scalar metric cannot capture. Private hands are filtered by reach probability to retain strategically relevant holdings, evaluated at $K \in \{20, 50, 100\}$ buckets. Each condition runs 10 independent seeds (each drawing a fresh feature-discovery sample), and we report the mean, standard deviation, and a 95\% confidence interval over the seeds.

\paragraph{ROVER Trials.}
To test whether the agent \emph{constructs} strategic features rather than retrieving memorized poker knowledge, we introduce \emph{ROVER Trials}, an original two-player zero-sum betting game of hidden information designed for this paper and absent from any pretraining corpus. Each player privately holds a \emph{rover}, described by three integer dials each in $\{0,\dots,4\}$: \textsc{Power} (raw engine strength), \textsc{Grip} (traction on rough ground), and \textsc{Affinity} (a single specialized terrain type), giving $5^3 = 125$ private types drawn uniformly and independently. Both players ante one chip and play two betting rounds; between them a public \emph{terrain} is revealed uniformly from five types (sand/grass are smooth; rock/mud/ice are rough). A rover travels distance $\mathrm{Power} + \mathrm{Grip}\cdot\mathbb{1}[\text{terrain rough}] + b\cdot\mathbb{1}[\textsc{Affinity} = \text{terrain}]$; the farther rover wins the pot. We set the affinity bonus $b = 10$, making a rover a ``specialist'' that is dominant on its own (one-in-five) terrain but ordinary elsewhere; expected distance alone then cannot separate a volatile specialist from a steady rover of the same mean. The LLM receives only the rules text, with no poker terminology, no feature hints, and no evaluator, so any useful abstraction must be assembled from the rules. We evaluate $K \in \{5, 10, 20, 30\}$ buckets over 10 independent seeds per backbone. Baselines mirror the poker setting: \emph{Rank} buckets rovers by expected distance (the ROVER analog of EHS) and requires the payoff simulator; \emph{PA} clusters the 2D potential-aware features $(\text{expected distance},\ \mathrm{std}_{\text{terrain}})$, the analog of potential-aware abstraction; \emph{Random} and \emph{full} (no abstraction) complete the comparison. Solving uses CFR+ for 500 iterations, with exploitability computed by exact best response on the unabstracted tree.

\paragraph{Cross-game settings.}
Three further settings test portability of the unchanged pipeline. In Pot-Limit Omaha (PLO4) preflop, the full hand space of $\binom{52}{4} = 270{,}725$ raw combinations reduces to 16,432 canonical representatives after suit isomorphism, clustered with $K=30$; no exploitability is computed, and we instead inspect whether the discovered features capture recognized PLO strategic concepts. In HUNL preflop (169 canonical hands, $K=10$) and flop (1,176 valid holdings, $K=20$), we test cross-street generalization without pipeline modification. Finally, to move beyond card games entirely, we apply the agent to Japanese Riichi Mahjong \citep{mahjong_repo}, a tile-based game with 136 tiles, chance nodes (random draws from the wall), and no shared structure with poker: we fix a mid-game scenario (turn 8, East round, 14 visible discards) and sample 200 random 13-tile hands ($K=10$).

\paragraph{Baselines.}
We compare against the standard information-abstraction baselines used in prior poker-solving work \citep{johanson2013measuring, ganzfried2014potential}, plus a random lower bound and an LLM text-embedding baseline. \emph{EHS} assigns each hand its expected equity against a uniform opponent range over all possible river cards and clusters this scalar with $K$-means; it requires a hand evaluator and Monte Carlo rollout. \emph{Potential-Aware (PA)} clusters hands in the 2D space of (EHS, $\sqrt{\mathrm{Var}[\text{HS}]}$), with variance computed across all river cards \citep{ganzfried2014potential}; it requires a hand evaluator plus full river enumeration. \emph{LLM-Embed (zero-shot)} has the LLM write a text description of each hand, embeds the descriptions with a sentence encoder, and clusters them with $K$-means, testing whether raw embeddings capture useful structure without explicit feature discovery. \emph{Random} assigns hands to buckets uniformly at random. Both evaluator-based baselines use domain-specific code that our agent does not.

\paragraph{Evaluation metric.}
For HUNL endgames, we report exploitability of the lifted strategy as a fraction of the pot. The lifted strategy is obtained by running CFR for 2,000 iterations on the abstract game and mapping the resulting strategy back to the original game. Lower exploitability indicates a better abstraction. We also report the \textbf{ratio} of agent exploitability to EHS exploitability; ratio $< 1$ means the agent outperforms EHS.

\paragraph{Implementation and cost.}
We use GPT-5.5 via an OpenAI-compatible API for the main results. For the turn subgames, the \texttt{GameConfig} supplies the public board, pot, and a summary of the players' range composition (nonzero-hand counts and entropies), all available at solving time without any hand evaluator or solver output; a rules-only variant that omits the range summary behaves identically (Appendix~\ref{app:prompt_robustness}). To test backbone dependence we run the same pipeline with open-weight Llama-3.3-70B and Qwen-2.5-32B. Feature discovery is a single LLM call per game at temperature $0.7$ with high reasoning effort; private-state scoring uses temperature $0.0$ with deterministic decoding and low reasoning effort, in batches of 20--30 holdings with 3-attempt retries. We define \emph{parse rate} as the fraction of scoring responses that match the required structured format without manual correction. Features are filtered (min std $= 0.10$, max correlation $= 0.90$) and rescaled before $K$-means++ clustering (10 restarts, 50 iterations each). CFR uses discounted updates (DCFR; $\alpha=1.5$, $\beta=0$, $\gamma=2$) with alternating regret matching for 2,000 iterations; exploitability is computed exactly by best-response traversal over the unabstracted game tree. Each seed redraws both the feature-discovery sample and the $K$-means initialization; prompts are identical across games except the \texttt{GameConfig} content (full prompts in Appendix~\ref{app:prompts}). Clustering stability across seeds yields mean pairwise ARI of 0.77. Per seed the pipeline makes one feature-discovery call and roughly ten scoring batches, on the order of a dozen LLM calls; wall-clock is dominated by API latency (a few to $\sim$15 minutes per seed), while local feature selection and clustering take under a second. Scoring depends only on the discovered features and holdings, not on $K$, so its cost is amortized across granularities: the $K{\in}\{20,50,100\}$ abstractions for a seed reuse a single scoring pass. Downstream CFR (2,000 DCFR iterations) is the dominant compute cost and grows with $K$.

%% file: sections/results.tex
\section{Results}

\paragraph{HUNL turn endgames.}
\label{sec:results}

Table~\ref{tab:turn-results} reports lifted-strategy exploitability on the two turn endgames. For each subgame and granularity $K$ we run the pipeline for 10 independent seeds (each with a fresh feature-discovery sample at temperature $0.7$) and report the mean and standard deviation of exploitability, its ratio to the EHS baseline with a $95\%$ confidence interval, and the best run; ratios below $1.0$ mean the agent outperforms EHS. All runs use GPT-5.5 and parse at $100\%$.

\begin{table}[t]
\centering
\small
\caption{HUNL turn endgame exploitability (fraction of pot; mean over 10 seeds). \textbf{Ratio} = mean Agent / EHS $\pm$ 95\% Student-$t$ CI half-width.}
\label{tab:turn-results}

\begin{tabular}{lrrrr}
\toprule
Board & $K$ & Agent & EHS & Ratio \\
\midrule
\multirow{3}{*}{7$\spadesuit$9$\heartsuit$9$\clubsuit$T$\clubsuit$}
 & 20  & 0.0437 & 0.0335 & 1.30 $\pm$ 0.20 \\
 & 50  & 0.0142 & 0.0192 & 0.74 $\pm$ 0.17 \\
 & 100 & 0.0036 & 0.0098 & \textbf{0.36} $\pm$ 0.04 \\
\midrule
\multirow{3}{*}{T$\spadesuit$6$\heartsuit$A$\heartsuit$7$\clubsuit$}
 & 20  & 0.1359 & 0.1088 & 1.25 $\pm$ 0.23 \\
 & 50  & 0.0483 & 0.0592 & 0.82 $\pm$ 0.10 \\
 & 100 & 0.0102 & 0.0256 & \textbf{0.40} $\pm$ 0.08 \\
\bottomrule
\end{tabular}
\end{table}

As $K$ grows from 20 to 100, the mean exploitability ratio falls from 1.28 to 0.78 to 0.38. At $K{=}100$ the agent cuts exploitability by roughly 62\% relative to EHS, with a 95\% CI below 1.0 on both subgames and all 20 runs beating EHS. Finer abstractions give the multi-dimensional feature space more room to separate hands, while a one-dimensional EHS ordering gains little from extra buckets once the major strength tiers are split. At $K{=}20$, by contrast, the agent underperforms EHS: with few buckets and after de-correlation filtering, the surviving feature basis is sometimes too thin to beat a clean equity ordering (Section~\ref{sec:analysis}).

\paragraph{Backbone dependence.}
If the gains come from the backbone's internalized strategic knowledge, weaker models under the same prompt should not reproduce them. Table~\ref{tab:backbone} confirms this: only the frontier GPT-5.5 reliably beats EHS; Llama-3.3-70B is near parity; Qwen-2.5-32B is consistently worse.

\begin{table}[t]
\centering
\small
\caption{Backbone dependence: mean exploitability ratio (Agent / EHS) over 10 seeds.}
\label{tab:backbone}
\begin{tabular}{llrr}
\toprule
$K$ & Backbone & SG1 & SG2 \\
\midrule
\multirow{3}{*}{50}
 & GPT-5.5            & 0.74 & 0.82 \\
 & Llama-3.3-70B      & 1.62 & 1.86 \\
 & Qwen-2.5-32B       & 1.48 & 2.01 \\
\midrule
\multirow{3}{*}{100}
 & GPT-5.5            & \textbf{0.36} & \textbf{0.40} \\
 & Llama-3.3-70B      & 1.07 & 1.36 \\
 & Qwen-2.5-32B       & 1.96 & 3.70 \\
\bottomrule
\end{tabular}
\end{table}

\paragraph{Comparison with Other Methods.}
Table~\ref{tab:comparison} compares the agent with baselines at $K{=}50$. It beats a random assignment, the zero-shot LLM-Embed baseline, and EHS, without any domain-specific evaluator. PA, which has access to exact equity calculators and full river enumeration, achieves the lowest exploitability; the agent closes about a third of that gap using only LLM-derived features. We compare directly against methods whose abstraction signal is computable here: Deep CFR \citep{brown2019deepcfr} and WEVA \citep{li2026weva} need CFR iterations on the unabstracted game, infeasible at HUNL turn scale and undefined for PLO/Mahjong without a solver, and potential-aware abstraction \citep{ganzfried2014potential} needs full river enumeration. Where a computable analog exists, namely EHS and PA in poker and the Rank and 2D potential-aware baselines in ROVER (Section~\ref{sec:rover}), we include it directly.

\begin{table}[t]
\centering
\small
\caption{Comparison of abstraction methods at $K{=}50$ (exploitability as fraction of pot; agent is the 10-seed mean).}
\label{tab:comparison}
\begin{tabular}{lrr}
\toprule
Method & SG1 & SG2 \\
\midrule
Random & 0.1314 & 0.3670 \\
LLM-Embed (zero-shot) & 0.0318 & 0.1847 \\
\textbf{Abstraction Agent (ours)} & \textbf{0.0142} & \textbf{0.0483} \\
\midrule
\multicolumn{3}{l}{\textit{Domain-specific methods (require hand evaluator):}} \\
EHS & 0.0192 & 0.0592 \\
PA (EHS + Variance) & 0.0053 & 0.0254 \\
\bottomrule
\end{tabular}
\end{table}

\paragraph{ROVER Trials: Construction Beyond Memorization.}
\label{sec:rover}

Table~\ref{tab:rover} reports lifted-strategy exploitability ratios on ROVER Trials, an original game absent from any pretraining corpus; the LLM receives only the rules text and no poker terminology, evaluator, or game-tree traversal. The two baselines are \emph{Rank}, the expected-distance (EHS) analog, and \emph{PA}, the potential-aware 2D analog; both require the payoff simulator, which the agent does not use. GPT-5.5 beats the scalar Rank baseline at every granularity, with all 95\% CIs below 1.0, and the margin widens with granularity (from $0.92$ at $K{=}5$ to $0.66$ at $K{=}30$). This is the same pattern as on HUNL turn: once buckets are plentiful, a multi-dimensional feature space separates types that a scalar ordering must merge, here volatile terrain specialists from steady rovers of the same expected distance. Against the stronger PA baseline the agent improves at the mean for $K\in\{10,20,30\}$ (ratios $0.91$--$0.94$), but those CIs overlap 1.0, so the advantage is directional: rules-only features roughly match, rather than clearly beat, a simulator-based 2D abstraction. The backbone gradient also mirrors HUNL: Llama-3.3-70B beats Rank at $K\ge10$ (ratio $0.76$ at $K{=}20,30$) but not PA, and Qwen-2.5-32B is below Rank only marginally and only at the finer granularities. The unabstracted full game has exploitability $0.0019$~chips, so abstraction still costs something; the relevant question is how much of the gap to evaluator-based abstractions a rules-only pipeline closes.

Because ROVER has no published strategy literature and no per-holding bucketing to copy, these results support the \emph{construction} reading: the LLM assembles useful features from the supplied rules rather than reciting a memorized recipe (Section~\ref{sec:analysis}).

\begin{table}[t]
\centering
\small
\caption{ROVER Trials: mean exploitability ratio (Agent\,/\,baseline) over 10 seeds $\pm$ 95\% Student-$t$ CI half-width.}
\label{tab:rover}
\begin{tabular}{llrr}
\toprule
Backbone & $K$ & Agent/Rank & Agent/PA \\
\midrule
\multirow{4}{*}{GPT-5.5}
 & 5  & 0.920 $\pm$ 0.074 & 1.043 $\pm$ 0.106 \\
 & 10 & 0.804 $\pm$ 0.038 & 0.940 $\pm$ 0.067 \\
 & 20 & 0.667 $\pm$ 0.070 & 0.913 $\pm$ 0.106 \\
 & 30 & 0.663 $\pm$ 0.065 & 0.916 $\pm$ 0.111 \\
\midrule
\multirow{4}{*}{Qwen-2.5-32B}
 & 5  & 1.157 $\pm$ 0.085 & 1.319 $\pm$ 0.160 \\
 & 10 & 0.969 $\pm$ 0.059 & 1.128 $\pm$ 0.054 \\
 & 20 & 0.902 $\pm$ 0.058 & 1.238 $\pm$ 0.117 \\
 & 30 & 0.965 $\pm$ 0.078 & 1.328 $\pm$ 0.117 \\
\midrule
\multirow{4}{*}{Llama-3.3-70B}
 & 5  & 1.102 $\pm$ 0.103 & 1.254 $\pm$ 0.162 \\
 & 10 & 0.924 $\pm$ 0.060 & 1.080 $\pm$ 0.088 \\
 & 20 & 0.760 $\pm$ 0.082 & 1.040 $\pm$ 0.116 \\
 & 30 & 0.760 $\pm$ 0.098 & 1.046 $\pm$ 0.147 \\
\bottomrule
\end{tabular}
\end{table}

\paragraph{Cross-game portability.}
Beyond HUNL turn, the same pipeline (unchanged code and prompts) yields coherent abstractions in PLO4 preflop (16,432 hands, $K{=}30$), HUNL flop (1,176 holdings, $K{=}20$), HUNL preflop (169 hands, $K{=}10$), and Riichi Mahjong (200 mid-game hands, $K{=}10$). The discovered features track each game's recognized strategy: suit coordination and straight connectivity in PLO4 \citep{billings2003approximating}, draw-versus-made separation on the flop, premium-pair grouping preflop, and shanten-speed / acceptance / draw-swing features in Mahjong, the last reflecting its chance nodes. Parse rates are near 100\%. Appendix~\ref{app:features} lists the features per game and summarizes the cross-game results (Table~\ref{tab:cross-game}); Appendix~\ref{app:plo_clusters}, \ref{app:mahjong_clusters}, \ref{app:flop_clusters}, \ref{app:preflop_clusters} show representative clusters.

These settings also mark the practical boundary of evaluator-based abstraction. EHS and PA presuppose a hand evaluator with Monte Carlo rollouts or full enumeration of future public cards, which becomes expensive for PLO4's 16,432 canonical four-card hands and is unavailable for Riichi Mahjong, where no equity evaluator or open-source solver exists. The agent's cost is instead linear in the number of holdings, about a dozen LLM calls per seed, which is what makes abstractions for these games obtainable here at all.


%% file: sections/analysis.tex
\section{Analysis}
\label{sec:analysis}

\paragraph{Why multi-feature abstraction outperforms EHS.}
EHS compresses all strategic information into a single scalar, losing vulnerability, nut potential, and blocker effects. The feature--EHS correlation analysis (Appendix~\ref{app:feature_ehs}) shows this: the made-strength feature correlates strongly with EHS ($r \approx 0.87$--$0.93$), while the improvement- and river-shift features have low or negative EHS correlation and thus capture complementary dimensions. Clustering in multi-dimensional space avoids merging hands with similar EHS but different strategic roles.

\paragraph{Granularity effects.}
The agent's advantage grows monotonically with granularity: the mean exploitability ratio falls from 1.28 at $K{=}20$ to 0.78 at $K{=}50$ to 0.38 at $K{=}100$ (Table~\ref{tab:turn-results}). Finer abstractions partition the multi-dimensional space more precisely, while one-dimensional EHS gains less from extra buckets once the major strength tiers are separated. Clustering is stable across seeds: mean pairwise ARI is 0.77 overall, with $K{=}50$ (0.77--0.88) more stable than $K{=}20$ (0.63--0.80). Leave-one-out ablations (Appendix~\ref{app:feature_ablation}) confirm that \textsc{Vulnerability\_To\_Outdraws} and \textsc{Blocker\_Value} carry the most unique clustering signal.

\paragraph{Comparison with potential-aware methods.}
Potential-aware abstractions \citep{ganzfried2014potential} extend EHS with future-card distributions; EVPA \citep{li2025evpa} learns neural features online during CFR. Our \textsc{Improvement\_Potential} and \textsc{Vulnerability\_To\_Outdraws} capture analogous information through natural-language reasoning. PA achieves lower exploitability (Table~\ref{tab:comparison}), since it has exact equity calculators; our value is in \emph{beating} the evaluator-based EHS baseline and closing about a third of the gap to PA \emph{without} any game-specific evaluator or game-tree traversal. This zero-shot setting is the realistic case for games without existing solver implementations: no open-source Mahjong CFR solver exists, yet the pipeline produces meaningful strategic clusters in under a minute.

\paragraph{Constructed versus retrieved features.}
Because HUNL is heavily represented in pretraining corpora, one might worry that the LLM is \emph{retrieving} memorized poker features rather than \emph{constructing} them from the rules. This distinction does not affect the practical contribution: the claim is that useful, evaluator-free abstractions can be elicited zero-shot from a rules description either way. Three observations support the construction reading. First, the same pipeline discovers chance-sensitive features in Riichi Mahjong 
and suit-coordination features in PLO4, domains with far thinner strategy literature than HUNL and no published per-hand bucketing to copy. Second, the backbone comparison (Table~\ref{tab:backbone}) shows that giving the \emph{same} rules and prompt to weaker models does not yield useful abstractions, so the gains track genuine strategic competence rather than surface recall. Third, and most directly, the ROVER Trials (Section~\ref{sec:rover}) are an original game introduced in this paper and absent from any pretraining corpus; the agent receives only its rules text, with no poker terminology, no evaluator, and no feature hints, yet the pipeline still discovers features that beat a scalar rank baseline and a 2D potential-aware baseline across most settings (Table~\ref{tab:rover}). Features that succeed on a game the model has never seen must be constructed from the rules rather than retrieved from memory.

%% file: sections/conclusion.tex
\section{Conclusion}

We presented an Abstraction Agent that uses LLMs to discover and score strategic features for information abstraction from natural-language game descriptions, without game-specific evaluators or game-tree traversal. On the two public HUNL turn endgames, the agent reduces exploitability relative to EHS by up to 62\%, with all $K{=}100$ runs beating EHS and the advantage growing in the bucket count. On ROVER Trials, an original game absent from any pretraining corpus, the rules-only pipeline beats the scalar rank baseline at every granularity, indicating construction from the rules rather than retrieval from memory; the unchanged pipeline also produces coherent abstractions in PLO4, HUNL preflop/flop, and Riichi Mahjong. These results position LLMs as structured interfaces that turn natural-language rules into formal computation.

%% file: sections/appendix.tex
\appendix

\section{Game Descriptions}
\label{app:games}

This appendix provides the complete game descriptions used as input to the Abstraction Agent. These descriptions constitute the \emph{only} domain knowledge provided to the system; no game-specific evaluation code, hand evaluators, or equity calculators are used.

\paragraph{Heads-Up No-Limit Texas Hold'em (HUNL).}
HUNL is played with a standard 52-card deck of 13 ranks ($2 < 3 < \cdots < K < A$) and 4 suits ($\spadesuit, \heartsuit, \diamondsuit, \clubsuit$). Each player receives 2 private (hole) cards, and five community cards are dealt face-up in stages: flop (3), turn (1), river (1). At showdown, each player forms the best 5-card combination from their 2 hole cards and 5 community cards (best 5 of 7). Hands are ranked, from strongest to weakest: straight flush (five consecutive ranks, all same suit), four of a kind, full house (three of one rank plus two of another), flush (five of one suit), straight (five consecutive ranks, mixed suits), three of a kind, two pair, one pair, and high card. Betting is no-limit: players may bet any amount up to their remaining stack, with a betting round after each community card stage. In our turn experiments, 4 community cards are visible and one more will be revealed, so hand rankings \emph{can change} with the final card: draws may complete, and currently strong hands may be overtaken.

\paragraph{Pot-Limit Omaha (PLO4).}
PLO4 uses the same 52-card deck, but each player receives \textbf{4 private cards} (compared to 2 in HUNL), with the same five community cards dealt in stages. The critical rule is that at showdown each player \textbf{must use exactly 2} of their 4 private cards combined with \textbf{exactly 3} of the 5 community cards; players cannot use 1, 3, or 4 private cards, which is the fundamental difference from HUNL. Hand rankings follow the same 9-category hierarchy (straight flush through high card). Betting is pot-limit: the maximum bet is the current pot size. In our PLO4 experiments, no community cards have been dealt, so hand value is entirely about \emph{post-flop potential}. The hand space is $\binom{52}{4} = 270{,}725$ raw combinations, reduced to 16,432 canonical representatives after suit isomorphism.

\paragraph{Riichi Mahjong (Japanese Competitive Mahjong).}
Riichi Mahjong is played with 136 tiles: 34 unique types $\times$ 4 copies, comprising three numbered suits of 1--9 (Man/Characters, Pin/Circles, Sou/Bamboo) plus honor tiles (4 winds and 3 dragons). A complete hand (14 tiles) consists of 4 sets plus 1 pair, where each set is either a sequence (three consecutive tiles of one suit) or a triplet (three identical tiles); Seven Pairs and Thirteen Orphans are special patterns. Progress toward completion is measured by the \emph{shanten} number, the minimum tile exchanges to reach \emph{tenpai} (shanten 0, one tile from winning). Each turn a player draws one tile randomly from the remaining wall ($\sim$70 tiles initially, $\sim$40--50 mid-game); the draw may complete the hand, reduce shanten, enable higher-value scoring patterns (yaku), or provide no improvement, and it is the primary source of randomness after the initial deal. Hand value depends on yaku such as Riichi, Tanyao, Pinfu, Honitsu, and Chinitsu; more han means exponentially higher payment, with mangan (5+ han) worth 8000 points. Strategically, a hand is characterized by progression speed (shanten plus acceptance width), value potential (yaku routes), tile efficiency, defensive safety (safe discard availability), and waiting pattern quality. We use the \texttt{mahjong} Python library \citep{mahjong_repo} for shanten calculation and hand evaluation.

\paragraph{ROVER Trials.}
ROVER Trials is an original two-player zero-sum betting game of hidden information designed for this paper. It appears in no pretraining corpus, so the LLM cannot retrieve memorized strategy knowledge from it. The agent receives only the rules text below, with no feature hints and no evaluator. Each player is privately and independently assigned one \emph{rover}, described by three integer dials each in $\{0,\dots,4\}$: \textsc{Power} (raw engine strength), \textsc{Grip} (traction on rough ground), and \textsc{Affinity} (the single terrain type the rover is specialized for), giving $5^3 = 125$ possible rovers; both players draw uniformly and independently, so they may even hold the same rover. Play proceeds as follows. Both players ante one chip. In betting round 1, before the terrain is known, the first player may check or raise (+2 chips); facing a raise a player may fold, call, or re-raise, with at most 2 raises per round. Then the \emph{race terrain} is revealed publicly: one of five terrain types, each equally likely; sand and grass are \emph{smooth}, while rock/mud/ice are \emph{rough}. Betting round 2 proceeds after the terrain is revealed (raise size +4 chips) with the same betting rules. If neither player folds, both rovers race and the one that travels farther wins the pot; equal distance splits it. A rover travels
\begin{align*}
\mathrm{distance} = \textsc{Power} &+ \textsc{Grip}\cdot\mathbb{1}[\text{terrain rough}] \\
&+ b\cdot\mathbb{1}[\textsc{Affinity} = \text{terrain}],
\end{align*}
where the affinity bonus is set to $b = 10$ in our experiments. A rover is therefore a ``specialist'' that is dominant on its own (one-in-five) terrain but ordinary elsewhere; expected distance alone cannot separate a volatile specialist from a steady rover of the same mean, which is precisely the regime in which a multi-feature abstraction can beat a scalar baseline. For evaluation we use $K \in \{5, 10, 20, 30\}$ buckets over 10 independent seeds per backbone (GPT-5.5, Llama-3.3-70B, Qwen-2.5-32B). Baselines mirror the poker setting: \emph{Rank} buckets rovers by expected distance (the ROVER analog of EHS) and requires the payoff simulator; \emph{PA} clusters the 2D potential-aware features $(\mathbb{E}[\mathrm{distance}],\ \mathrm{std}_{\text{terrain}})$; \emph{Random} assigns buckets uniformly; and \emph{full} is the unabstracted game. Solving uses CFR+ for 500 iterations; exploitability is computed exactly by best response on the unabstracted tree. The unabstracted full game reaches $0.0019$~chips of exploitability, so the abstraction still incurs a cost; the relevant question is how much of the gap to evaluator-based abstractions a rules-only pipeline closes.

\section{PLO4 Preflop Hand Taxonomy}
\label{app:plo_taxonomy}

PLO4 preflop hands are characterized along multiple strategic dimensions; suit structure in particular determines flush potential. Table~\ref{tab:plo_suit_patterns} and Table~\ref{tab:plo_structure_types} provide a taxonomy.


\begin{table}[h]
\centering
\small
\caption{PLO4 suit pattern classification (canonical hand counts out of 16,432).}
\label{tab:plo_suit_patterns}
\begin{tabular}{@{}lcp{3.0cm}@{}}
\toprule
Pattern & Count & Description \\
\midrule
Double-suited (2+2) & 3,081 & Two suited pairs, two flush draws. \\
Single-suited (2+1+1) & 7,098 & One suited pair; one flush draw. \\
Triple-suited (3+1) & 3,718 & Three of one suit, one weaker flush draw. \\
Rainbow (1+1+1+1) & 1,820 & All four suits differ; no flush potential. \\
Monotone (4+0) & 715 & Same suit; one flush draw (must-use-two). \\
\bottomrule
\end{tabular}
\end{table}


\begin{table*}[h]
\centering
\small
\caption{PLO4 structural hand types with examples and strategic properties.}
\label{tab:plo_structure_types}

\begin{tabular}{lll}
\toprule
Type & Example & Strategic Properties \\
\midrule
Premium pairs + suited & AA-KK ds & Top pair + flush draws; strongest \\
Top rundowns & AKQJ, KQJT & Max straight connectivity \\
Suited connectors & 8h7h6s5s & Flush + straight; strong playability \\
High pairs + danglers & AA-xx rainbow & High pair, limited coordination \\
Mid rundowns & T987, 9876 & Good connectivity, lower nut potential \\
Suited aces & As-xxx & Nut flush draw potential \\
Trips & AAA-x & Reduced pair potential; weak in PLO \\
Low connected & 5432, 6543 & Straight potential but rarely nut \\
Disconnected low & 7h3d2c9s & No coordination; weakest \\
\bottomrule
\end{tabular}
\end{table*}


\paragraph{Key PLO4 Concepts.}
Four concepts drive PLO4 preflop evaluation. \emph{Nut potential}: because multi-way pots are common, hands that can make the \emph{nut} flush (holding the Ace of a suit) or nut straight are far more valuable than hands that can only make non-nut versions. \emph{Card coordination}: all 4 cards should ``work together,'' contributing to the same straights or flushes; a hand like $A\spadesuit K\spadesuit Q\heartsuit J\heartsuit$ (double-suited, connected) has maximum coordination, while $A\spadesuit 7\diamondsuit 3\clubsuit 2\heartsuit$ has a ``dangler'' (disconnected card) that reduces effective hand strength. \emph{Domination risk}: hands whose flush or straight draws can only make non-nut hands may hit their draw and still lose to a higher version; small pairs without backup draws are particularly vulnerable. \emph{The must-use-two rule}: using exactly 2 private cards fundamentally changes hand evaluation compared to HUNL; holding four cards of one suit does \emph{not} help make a flush (only 2 can be used), and holding trips reduces pair potential (only 1 card of that rank remains in the deck).

\section{Prompt Templates}
\label{app:prompts}

This section provides the exact prompt templates used by the Abstraction Agent. All prompts are game-agnostic; domain knowledge enters only through the \texttt{GameConfig} fields.

\paragraph{Feature Discovery Prompt.}

\begin{small}
\begin{verbatim}
SYSTEM:
You are a game-theory analyst specializing in
imperfect-information games. Your task is to
propose continuous strategic features for
grouping private holdings.

USER:
## Game Description
{config.game_description}

## Task
Propose {M} continuous features in [0, 1] for
evaluating private holdings in this game.

## Requirements
1. SPREAD: Each feature must produce sufficient
   variance (std >= 0.15) across the hand space.
2. INDEPENDENCE: Features should capture different
   strategic dimensions (pairwise |r| < 0.7).
3. COMPLETENESS: Together, features should cover
   the major axes of strategic variation.
4. CALIBRATION: Provide anchors mapping 0.0, 0.5,
   and 1.0 to concrete game situations.
{5. FUTURE AWARENESS (if has_future_cards):
   Features should capture both current state AND
   potential for change when new information is
   revealed. Note: {future_card_note}}

## Anti-Patterns (DO NOT propose)
- Binary features (e.g., "has a pair: yes/no")
- Features that are subsets of another
- Features requiring external computation

## Output Format (JSON)
[{"name": "FEATURE_NAME",
  "description": "what it measures",
  "anchors": {"0.0": "...", "0.5": "...",
              "1.0": "..."}}]
\end{verbatim}
\end{small}

\paragraph{Hand Scoring Prompt.}

\begin{small}
\begin{verbatim}
SYSTEM:
{config.game_context}

You will score private holdings on the following
features. Use the calibration anchors as reference
points for consistent scoring.

Features:
{for each feature f:
  - {f.name}: {f.description}
    0.0 = {f.anchors["0.0"]}
    0.5 = {f.anchors["0.5"]}
    1.0 = {f.anchors["1.0"]}}

USER:
{board_to_text(board)}

Score each holding below. Output format:
INDEX: FEAT1=0.xx FEAT2=0.yy ...

{for i, hand in enumerate(batch):
  {i}: {hand_to_text(hand, board)}}
\end{verbatim}
\end{small}

\paragraph{Example Scoring Output.}

For a HUNL turn board $7\spadesuit\, 9\heartsuit\, 9\clubsuit\, T\clubsuit$:

\begin{small}
\begin{verbatim}
0: CURRENT_MADE_STRENGTH=0.85
   IMPROVEMENT_POTENTIAL=0.30
   VULNERABILITY=0.45 NUT_CEILING=0.90
   BLOCKER_VALUE=0.60 RIVER_STABILITY=0.70
1: CURRENT_MADE_STRENGTH=0.20
   IMPROVEMENT_POTENTIAL=0.80
   VULNERABILITY=0.15 NUT_CEILING=0.40
   BLOCKER_VALUE=0.25 RIVER_STABILITY=0.20
...
\end{verbatim}
\end{small}

\section{Experimental Details}
\label{app:exp_details}

\paragraph{Libratus Subgame Specifications.}

Table~\ref{tab:subgame_specs} provides details of the two Libratus-derived turn subgames used in our quantitative HUNL evaluation. Both are turn boards (4 community cards, 1 more to come), and hands are filtered by reach probability.

\begin{table}[h]
\centering
\small
\caption{Libratus subgame specifications.}
\label{tab:subgame_specs}
\begin{tabular}{clcc}
\toprule
ID & Board & Street & Pot \\
\midrule
1 & $7\spadesuit\, 9\heartsuit\, 9\clubsuit\, T\clubsuit$ & Turn & 500 \\
2 & $T\spadesuit\, 6\heartsuit\, A\heartsuit\, 7\clubsuit$ & Turn & 4780 \\
\bottomrule
\end{tabular}
\end{table}

\paragraph{Board Texture Analysis.}
Subgame 1 ($7\spadesuit\, 9\heartsuit\, 9\clubsuit\, T\clubsuit$) is a paired board with a club flush draw and open-ended straight draws; the pair creates full house potential, and the connected texture enables many draws. Subgame 2 ($T\spadesuit\, 6\heartsuit\, A\heartsuit\, 7\clubsuit$) is an ace-high board with a heart flush draw, where the disconnected low cards create a wide range of possible hand strengths.

\paragraph{CFR Solver Configuration.}

\begin{itemize}
    \item Algorithm: Discounted CFR (DCFR; $\alpha=1.5$, $\beta=0$, $\gamma=2$) with alternating updates
    \item Iterations: 2,000
    \item Exploitability: Exact best-response computation over the unabstracted game tree
    \item Lifted exploitability: Strategy from abstract game mapped back to original game, then best-response computed
\end{itemize}

\paragraph{LLM API Configuration.}

\begin{itemize}
    \item Model: GPT-5.5 (primary), Llama-3.3-70B and Qwen-2.5-32B (comparison)
    \item Temperature: 0.7 (feature discovery), 0.0 (scoring, deterministic)
    \item Batch size: 20--30 hands per API call
    \item Retry logic: 3 attempts with exponential backoff
    \item Caching: All LLM responses cached by (model, prompt hash) for reproducibility
\end{itemize}

\section{Discovered Features by Game}
\label{app:features}

Table~\ref{tab:cross-game} summarizes all six evaluation settings; exploitability is measured on HUNL turn and ROVER, and the remaining settings are evaluated by inspecting the discovered features and clusters. The rest of this section lists the features the agent discovers in the HUNL turn and PLO4 preflop settings, with their calibration anchors.

\begin{table}[h]
\centering
\small
\caption{Cross-game summary of the six evaluation settings.}
\label{tab:cross-game}

\begin{tabular}{@{}lrll@{}}
\toprule
Game & $N$ & $K$ & Key Result \\
\midrule
HUNL Turn & ${\sim}$1K & 20--100 & up to 62\% $\downarrow$ vs EHS \\
ROVER Trials & 125 & 5--30 & beats baseline \\
PLO4 Preflop & 16,432 & 30 & PLO features \\
HUNL Flop & 1,176 & 20 & Draw/made sep. \\
HUNL Preflop & 169 & 10 & Premium groups \\
Mahjong & 200 & 10 & Chance features \\
\bottomrule
\end{tabular}%

\end{table}

\paragraph{HUNL Turn Features.}

The following features are consistently discovered across turn subgames:

\begin{enumerate}
    \item \textbf{CURRENT\_MADE\_STRENGTH} (retained in both subgames)\\
    Measures the current 5-card hand ranking relative to all possible holdings.\\
    Anchors: 0.0 = no pair/weak high card; 0.5 = top pair/overpair; 1.0 = set or better.

    \item \textbf{IMPROVEMENT\_POTENTIAL} (retained in both subgames)\\
    Probability of improving to a significantly stronger hand with the river card.\\
    Anchors: 0.0 = no draws; 0.5 = gutshot or weak draw; 1.0 = open-ended straight + flush draw.

    \item \textbf{VULNERABILITY\_TO\_OUTDRAWS} (retained in both subgames)\\
    Risk that currently strong hands will be overtaken by opponent draws.\\
    Anchors: 0.0 = invulnerable (nuts); 0.5 = moderate risk; 1.0 = highly vulnerable to multiple draws.

    \item \textbf{NUT\_CEILING} (retained in both subgames)\\
    How close the hand is to the best possible hand on this board.\\
    Anchors: 0.0 = far from nuts; 0.5 = second/third nuts; 1.0 = current nuts.

    \item \textbf{BLOCKER\_VALUE} (retained in both subgames)\\
    Whether the hand blocks opponent's strong holdings or draws.\\
    Anchors: 0.0 = no blocking; 0.5 = blocks one draw; 1.0 = blocks multiple strong hands.

    \item \textbf{RIVER\_STABILITY} (retained in 1/2 subgames)\\
    How stable the hand's relative strength is across possible river cards.\\
    Anchors: 0.0 = highly volatile; 0.5 = moderate stability; 1.0 = rank unchanged by any river.

    \item \textbf{RIVER\_EQUITY} (sometimes filtered)\\
    Expected equity after the river card is revealed.\\
    Often filtered due to high correlation with CURRENT\_MADE\_STRENGTH.
\end{enumerate}

\paragraph{PLO4 Preflop Features.}

\begin{enumerate}
    \item \textbf{RANK\_AND\_PAIR\_STRENGTH} (retained)\\
    High-card value and pair potential of the 4-card holding.\\
    Anchors: 0.0 = low unpaired disconnected; 0.5 = medium pair or high-card; 1.0 = AA/KK coordinated.

    \item \textbf{SUITEDNESS\_AND\_\allowbreak FLUSH\_POTENTIAL} (retained)\\
    Suit structure and nut flush draw potential.\\
    Anchors: 0.0 = rainbow; 0.5 = single-suited; 1.0 = double-suited, ideally ace-high.

    \item \textbf{STRAIGHT\_CONNECTIVITY} (retained)\\
    How many different straights the hand can make.\\
    Anchors: 0.0 = widely separated ranks; 0.5 = one or two gaps; 1.0 = 4-card rundown.

    \item \textbf{ALL\_CARD\_COORDINATION} (retained)\\
    Degree to which all 4 cards work together.\\
    Anchors: 0.0 = severe danglers; 0.5 = 3 coordinated; 1.0 = all 4 coordinated.

    \item \textbf{NUT\_POTENTIAL} (retained)\\
    How often the holding makes the best (nut) version of strong hands rather than dominated second-best hands.\\
    Anchors: 0.0 = mostly weak or dominated draws; 0.5 = some nut-relevant components; 1.0 = excellent nut-making capacity.
\end{enumerate}

\textsc{Flop\_Value\_Shift\_Potential} and \textsc{Domination\_Risk\_Resistance} were also discovered but dropped by the correlation filter ($|r| = 0.91$ with \textsc{All\_Card\_Coordination} and $|r| = 0.97$ with \textsc{Nut\_Potential}, respectively).

\section{HUNL Preflop Cluster Details}
\label{app:preflop_clusters}

Table~\ref{tab:preflop-clusters-full} shows the complete clustering of 169 canonical HUNL preflop hands into $K=10$ clusters. The agent discovers 7 features, all retained after filtering: \textsc{High\_Card\_Power}, \textsc{Pair\_And\_Set\_Value}, \textsc{Suited\_Flush\_Potential}, \textsc{Straight\_Connectivity}, \textsc{Domination\_Resilience}, \textsc{Equity\_Realization\_Stability}, and \textsc{Runout\_Volatility}. Parse rate is 100\%. The clusters separate premium pairs, suited connectors, offsuit broadways, and junk hands without any poker-specific code.

\begin{table*}[h]
\centering
\small
\caption{Complete HUNL preflop clusters ($K=10$, 169 canonical hands, sorted by ID).}
\label{tab:preflop-clusters-full}
\begin{tabular}{cll}
\toprule
Cluster & Representative Hands & Characteristics \\
\midrule
0 (8) & 22, 33, 44, 55, 66, 77, 88, 99 & Small/medium pairs (set-mining value) \\
1 (19) & A2s--A5s, 54s--76s & Suited aces + suited connectors \\
2 (19) & 32o--T2o & Offsuit junk (weakest hands) \\
3 (12) & QTo, KTo, ATo, QJo, KJo, AJo & Offsuit broadways \\
4 (25) & K2s--K4s, Q2s--Q4s, T4s--J4s & Suited high + low kicker \\
5 (10) & ATs, QJs, KJs, AJs, KQs & Premium suited broadways \\
6 (23) & 32s--92s, T2s & Suited junk (low cards) \\
7 (18) & A3o--A5o, 54o--76o & Offsuit aces + connectors \\
8 (6) & TT, JJ, QQ, KK, AKo, AA & Premium pairs + AKo \\
9 (29) & J2o--K3o, A2o & Offsuit high + low kicker \\
\bottomrule
\end{tabular}
\end{table*}

The clustering shows strategic coherence: Cluster~8 isolates the strongest starting hands (premium pairs and AKo), Cluster~5 groups premium suited broadways with flush and straight potential, and Cluster~0 groups all small-to-medium pairs together (sharing similar set-mining value). The agent correctly separates suited from offsuit hands (Clusters 1/4/6 vs 7/9/2), reflecting the importance of flush potential in preflop hand evaluation.

\section{Feature--EHS Correlation}
\label{app:feature_ehs}

Table~\ref{tab:feature-ehs} reports the Pearson correlation between each discovered feature and EHS on the two turn subgames. The made-strength feature correlates strongly with EHS ($r \approx 0.87$--$0.93$), confirming that the agent recovers the standard strength signal, while the improvement- and river-shift features have low or negative correlation and thus capture strategic dimensions that a scalar EHS ordering cannot represent.

\begin{table}[h]
\centering
\small
\caption{Pearson correlation between LLM features and EHS across turn subgames.}
\label{tab:feature-ehs}
\begin{tabular}{@{}lrr@{}}
\toprule
Feature & SG1 & SG2 \\
\midrule
Current\_Made\_Str. & 0.87 & 0.93 \\
Improvement\_Pot. & 0.27 & $-$0.30 \\
Vuln.\_To\_Outdraws & 0.68 & 0.31 \\
Nut\_Ceiling & 0.68 & 0.14 \\
River\_Stability & $-$0.36 & 0.52 \\
Blocker\_Value & 0.65 & 0.46 \\
\bottomrule
\end{tabular}
\end{table}

\section{PLO4 Preflop Cluster Examples}
\label{app:plo_clusters}

Tables~\ref{tab:plo-clusters} and~\ref{tab:plo-clusters-cont} show the complete clustering of the 16,432 canonical PLO4 preflop hands into $K=30$ clusters; the clusters separate hands by rank strength, suitedness, connectivity, and coordination without poker-specific code. Since PLO4 preflop hands are suit-isomorphic (only the \emph{pattern} of shared suits matters, not which specific suits are held), we use abstract suit labels $a, b, c, d$ assigned by first-appearance order. The five canonical suit patterns are:
\textbf{ds} (double-suited, 2+2): two pairs share suits, e.g., A$_a$K$_a$Q$_b$J$_b$;
\textbf{ss} (single-suited, 2+1+1): one pair shares a suit, e.g., A$_a$K$_a$Q$_b$J$_c$;
\textbf{r} (rainbow, 1+1+1+1): all four suits differ, e.g., A$_a$K$_b$Q$_c$J$_d$;
\textbf{mono} (monotone, 4+0): all same suit, e.g., A$_a$K$_a$Q$_a$J$_a$;
\textbf{ts} (triple-suited, 3+1): three share a suit, e.g., A$_a$K$_a$Q$_a$J$_b$.

\begin{table*}[h]
\centering
\small
\caption{Complete PLO4 preflop clusters ($K=30$, 16,432 canonical hands; abstract suit labels $a$--$d$).}
\label{tab:plo-clusters}
\begin{tabular}{rp{5.5cm}l}
\toprule
Cluster & Representative Hands & Characteristics \\
\midrule
0 (590) & 9$_a$7$_b$5$_a$5$_b$ (ds), 6$_a$6$_b$5$_a$3$_b$ (ds) & Low-rank suited coordination holdings \\
1 (505) & A$_a$J$_b$J$_c$7$_b$ (ss), A$_a$J$_b$J$_c$6$_b$ (ss) & High-pair heavy \\
2 (629) & K$_a$T$_b$9$_a$5$_b$ (ds), A$_a$T$_a$5$_b$2$_b$ (ds) & Low-rank suited coordination holdings \\
3 (324) & A$_a$A$_b$K$_c$5$_c$ (ss), A$_a$A$_b$9$_c$7$_c$ (ss) & Premium paired / AA-heavy holdings \\
4 (743) & Q$_a$9$_a$6$_b$3$_b$ (ds), J$_a$T$_b$7$_a$3$_a$ (ts) & Low-rank suited coordination holdings \\
5 (468) & Q$_a$T$_b$9$_c$6$_a$ (ss), Q$_a$J$_b$8$_a$8$_c$ (ss) & Mixed suited coordination holdings \\
6 (168) & A$_a$A$_b$5$_a$4$_b$ (ds), A$_a$A$_b$J$_a$J$_b$ (ds) & Premium paired / AA-heavy holdings \\
7 (194) & K$_a$J$_b$9$_c$9$_d$ (r), A$_a$K$_b$T$_c$4$_d$ (r) & Rainbow high-card holdings \\
8 (702) & A$_a$9$_b$7$_c$4$_a$ (ss), A$_a$T$_b$T$_c$4$_a$ (ss) & Mixed suited coordination holdings \\
9 (482) & T$_a$T$_b$T$_c$2$_d$ (r), K$_a$Q$_b$4$_c$2$_d$ (r) & Medium-pair heavy \\
10 (314) & 5$_a$4$_a$4$_b$3$_a$ (ts), 7$_a$6$_b$4$_c$3$_a$ (ss) & Suited connected rundown holdings \\
11 (436) & J$_a$8$_b$8$_c$2$_d$ (r), Q$_a$6$_b$6$_c$4$_d$ (r) & Low-pair heavy \\
12 (496) & 7$_a$6$_a$4$_b$2$_a$ (ts), 9$_a$8$_a$5$_b$4$_a$ (ts) & Low-rank suited coordination holdings \\
13 (685) & Q$_a$T$_b$7$_b$6$_b$ (ts), Q$_a$J$_a$7$_a$5$_b$ (ts) & Low-rank suited coordination holdings \\
14 (401) & A$_a$5$_a$5$_b$4$_b$ (ds), J$_a$T$_b$9$_b$5$_a$ (ds) & Mixed suited coordination holdings \\
15 (889) & Q$_a$9$_a$9$_b$9$_c$ (ss), A$_a$J$_b$7$_b$5$_c$ (ss) & Unpaired \\
\bottomrule
\end{tabular}
\end{table*}

\begin{table*}[h]
\centering
\small
\caption{PLO4 preflop clusters (continued).}
\label{tab:plo-clusters-cont}
\begin{tabular}{rp{5.5cm}l}
\toprule
Cluster & Representative Hands & Characteristics \\
\midrule
16 (133) & J$_a$9$_b$7$_c$5$_d$ (r), Q$_a$T$_b$9$_c$7$_d$ (r) & Rainbow low-to-mid coordination holdings \\
17 (178) & K$_a$K$_b$T$_c$4$_d$ (r), A$_a$A$_b$Q$_c$7$_d$ (r) & Premium paired / AA-heavy holdings \\
18 (624) & Q$_a$J$_a$8$_b$2$_a$ (ts), 5$_a$5$_b$2$_a$2$_b$ (ds) & Low-rank suited coordination holdings \\
19 (387) & 7$_a$5$_b$2$_c$2$_d$ (r), A$_a$7$_b$3$_c$2$_d$ (r) & Rainbow low-to-mid coordination holdings \\
20 (896) & 9$_a$7$_a$2$_a$2$_b$ (ts), Q$_a$7$_a$7$_b$2$_a$ (ts) & Low-pair heavy \\
21 (618) & A$_a$K$_b$9$_a$9$_b$ (ds), K$_a$Q$_b$T$_b$2$_a$ (ds) & Mixed suited coordination holdings \\
22 (661) & A$_a$K$_a$4$_b$3$_a$ (ts), A$_a$K$_b$T$_b$5$_b$ (ts) & Suited high-card coordination holdings \\
23 (981) & 9$_a$8$_b$5$_b$3$_c$ (ss), Q$_a$9$_a$8$_b$2$_a$ (ts) & Low-rank suited coordination holdings \\
24 (275) & 9$_a$8$_a$6$_b$5$_b$ (ds), T$_a$9$_a$8$_a$6$_b$ (ts) & Suited connected rundown holdings \\
25 (213) & A$_a$K$_b$Q$_a$T$_a$ (ts), A$_a$K$_b$J$_b$T$_b$ (ts) & Suited connected rundown holdings \\
26 (1163) & J$_a$8$_a$6$_a$3$_b$ (ts), A$_a$9$_b$7$_b$2$_c$ (ss) & Low-rank suited coordination holdings \\
27 (354) & A$_a$K$_b$J$_b$4$_a$ (ds), A$_a$Q$_b$J$_b$9$_b$ (ts) & Suited high-card coordination holdings \\
28 (826) & 7$_a$4$_b$3$_a$3$_c$ (ss), T$_a$6$_b$4$_c$3$_c$ (ss) & Low-rank suited coordination holdings \\
29 (1097) & Q$_a$Q$_b$T$_c$7$_a$ (ss), K$_a$J$_b$9$_c$3$_c$ (ss) & Mixed suited coordination holdings \\
\bottomrule
\end{tabular}
\end{table*}

Cluster~17 contains exclusively rainbow holdings with premium pair strength (trips/quads such as AAA$+$ and high pairs), while Cluster~10 groups low-card suited rundowns with maximum straight connectivity and near-zero high-card content, demonstrating that the LLM-discovered features capture the multi-dimensional nature of PLO hand evaluation. Cluster~25 (top rundowns like AKQJ) and Cluster~10 (low rundowns like 9876) are cleanly separated despite both having very high straight connectivity, differing along the rank-strength and nut-potential dimensions.

\section{HUNL Flop Cluster Examples}
\label{app:flop_clusters}

Table~\ref{tab:flop-clusters} shows the complete clustering of the 1,176 flop holdings on the board A$\spadesuit$\,K$\heartsuit$\,Q$\diamondsuit$ into $K=20$ clusters. The agent cleanly separates made-strength tiers: sets in Cluster~5, the nut Broadway straight in Cluster~1, top two pair in Cluster~10, and top-pair hands split by kicker strength; draw-oriented holdings and no-interaction air form their own groups.

\begin{table*}[h]
\centering
\small
\caption{Complete HUNL flop clusters (board A$\spadesuit$ K$\heartsuit$ Q$\diamondsuit$; $K=20$, 1,176 hands, sorted by ID).}
\label{tab:flop-clusters}
\begin{tabular}{cll}
\toprule
Cluster & Representative Hands & Characteristics \\
\midrule
0 (80) & Ad 9s, Ad 9d, Ad 8c & Top pair A, strong kicker \\
1 (16) & Js Ts, Js Td, Jd Tc & Nut straight (Broadway) \\
2 (162) & 9s 4d, 9s 3c, 9s 2h & Air (9x low kicker, no draw) \\
3 (138) & Ah 5h, Ah 4h, Ks 9s & Medium made + backdoor draws \\
4 (25) & 9d 9c, 8s 8h, 6s 6d & Underpairs (below board) \\
5 (8) & Ad Ac, Ks Kd, Qs Qc & Sets (AA/KK/QQ) \\
6 (86) & 8c 7s, 8c 6d, 8c 5c & Junk (8x and below, no interaction) \\
7 (29) & Td 5d, Td 4s, Td 3c & T-kicker, weak draws \\
8 (153) & Ks 9d, Ks 8c, Ks 7d & Second pair K, weak kicker \\
9 (33) & Ks 4d, Ks 3s, Ks 2d & Second pair K, very weak kicker \\
10 (21) & Ad Ks, Ad Qh, Ac Kd & Top two pair (AK, AQ) \\
11 (35) & Ac 9s, Ac 9d, Ac 8c & Top pair A + backdoor nut flush \\
12 (23) & 9s 9d, 9s 9c, 8s 8d & Medium pocket pairs (multi-suit) \\
13 (39) & 8d 6s, 8d 5c, 8d 4s & Air + faint backdoor diamond \\
14 (43) & Jc 9s, Jc 8d, Jc 7c & J-kicker, straight potential \\
15 (35) & Ks Js, Ks Jd, Ks Ts & Second pair K + strong kicker J/T \\
16 (161) & 9s 8s, 9s 7d, 9s 6c & Largest junk group (below T) \\
17 (23) & Ts 3c, Ts 2d, Td 9s & T-kicker mixed, bottom straight draw \\
18 (27) & Ad Js, Ad Td, Ad Tc & Top pair A + strong kicker J/T (diamond) \\
19 (39) & Jd 7d, Jd 6s, Jd 5c & J-kicker + weak backdoor \\
\bottomrule
\end{tabular}
\end{table*}

\section{Prompt-Variant and Rules-Only Robustness}
\label{app:prompt_robustness}

Table~\ref{tab:prompt-robustness} reports the mean exploitability ratio (Agent\,/\,EHS, 10 seeds, GPT-5.5) for three feature-discovery prompt variants: \emph{rules-only} (the bare game rules), \emph{range-aware} (rules plus the public range summary; used in the main results), and \emph{feature-discovery} (range-aware plus an explicit instruction to separate made strength, draws, blockers, and range interaction). At $K{\in}\{20,50\}$ the three variants agree within the per-seed standard deviation (0.10--0.35), so the reported gains are not an artifact of a single prompt wording; in particular, the rules-only variant matches the range-aware one, indicating that the public range summary is not the source of the advantage.

\begin{table}[h]
\centering
\small
\caption{Prompt-variant robustness: mean Agent\,/\,EHS exploitability ratio over 10 seeds (GPT-5.5).}
\label{tab:prompt-robustness}
\begin{tabular}{lcccc}
\toprule
\multirow{2}{*}{Variant} & \multicolumn{2}{c}{$K{=}20$} & \multicolumn{2}{c}{$K{=}50$} \\
\cmidrule(lr){2-3}\cmidrule(lr){4-5}
 & SG1 & SG2 & SG1 & SG2 \\
\midrule
Rules-only        & 1.20 & 1.21 & 0.73 & 0.91 \\
Range-aware       & 1.30 & 1.25 & 0.74 & 0.82 \\
Feature-discovery & 1.35 & 1.18 & 0.76 & 0.92 \\
\bottomrule
\end{tabular}
\end{table}

\section{Feature Ablation Results}
\label{app:feature_ablation}

Table~\ref{tab:feature-ablation} reports leave-one-out feature importance, measured as the mean adjusted Rand index (ARI) between the clustering with and without each feature: a lower ARI means removing the feature changes the clustering more. Values come from a representative rules-only run, averaged across all subgame/$K$ conditions where the feature is selected. \textsc{Vulnerability\_To\_Outdraws} and \textsc{Blocker\_Value} carry the most unique clustering signal, while \textsc{River\_Equity} is largely redundant, consistent with its frequent removal by the correlation filter.

\begin{table}[h]
\centering
\small
\caption{Feature importance via leave-one-out ablation (mean ARI; lower = larger impact).}
\label{tab:feature-ablation}
\begin{tabular}{lr}
\toprule
Feature Removed & Mean ARI \\
\midrule
Vulnerability\_To\_Outdraws & 0.7558 \\
Blocker\_Value & 0.7607 \\
Current\_Made\_Strength & 0.7638 \\
Nut\_Ceiling & 0.7716 \\
Improvement\_Potential & 0.7866 \\
River\_Stability & 0.7959 \\
River\_Equity & 0.8122 \\
\bottomrule
\end{tabular}
\end{table}

\section{Riichi Mahjong Cluster Details}
\label{app:mahjong_clusters}

Table~\ref{tab:mahjong-features} shows the features discovered by the agent for Riichi Mahjong mid-game clustering: 6 discovered, 5 selected after correlation filtering. Table~\ref{tab:mahjong-clusters} shows representative clusters (100\% parse rate); the clusters separate hands by strategic role: offensive speed vs.\ defensive resilience vs.\ value potential.

\begin{table}[h]
\centering
\footnotesize
\setlength{\tabcolsep}{4pt}
\caption{Features discovered for Riichi Mahjong mid-game (turn 8, East round).}
\label{tab:mahjong-features}
\begin{tabular}{p{3.6cm}p{2.8cm}}
\toprule
Feature & Description \\
\midrule
\textsc{Current\_Speed} & Shanten + effective tiles \\
\textsc{Acceptance\_Breadth} & Tile types improving hand \\
\textsc{Hand\_Value\_Potential} & Scoring upside (yaku, dora) \\
\textsc{Wait\_Quality} & Quality of the winning wait \\
\textsc{Defensive\_Resilience} & Avoid dealing into opponents \\
\textsc{Next\_Draw\_Swing} & How one draw changes value (not selected) \\
\bottomrule
\end{tabular}
\end{table}

\begin{table}[h]
\centering
\footnotesize
\setlength{\tabcolsep}{4pt}
\caption{Complete Riichi Mahjong clusters ($K=10$, 200 hands).}
\label{tab:mahjong-clusters}
\begin{tabular}{rp{3.2cm}p{2.8cm}}
\toprule
Size & Representative Hands & Dominant Trait \\
\midrule
27 & 11578m 156p 13347s GR & Wide acceptance, 3--4 shanten \\
20 & 112467899m 4p 16s W & High speed + value \\
22 & 11149m 5579p 668s WH & Slow, high defense \\
16 & 1129m 39p 4777s WGR & High value, slow \\
23 & 1136m 2578p 1455s S & Average hands \\
33 & 1119m 1p 157889s WR & Most defensive \\
14 & 11378m 5688p 2s GRR & Most value potential \\
23 & 1239m 1245p 2667s W & Straight-type acceptance \\
15 & 12467m 489p 12339s & Balanced moderate \\
7 & 12456m 356p 56s WWR & Fastest, wide waits \\
\bottomrule
\end{tabular}
\end{table}

\section{Additional Implementation and Analysis Details}
\label{app:additional}

\paragraph{Game-Agnostic Design.}

The pipeline's game-agnosticism comes from a \texttt{GameConfig} interface specifying the game rules (for feature discovery), a shorter context (for scoring prompts), and two rendering callbacks (\texttt{hand\_to\_text}, \texttt{board\_to\_text}). To apply the agent to a new game, one writes a \texttt{GameConfig} (typically 20--50 lines of natural language) and two rendering functions. No evaluation code, rollout simulators, or domain-specific feature engineering is required.

\paragraph{Prompt Design Principles.}

Three design choices underlie the pipeline's output reliability: (1) \emph{Calibration anchors} map 0.0/0.5/1.0 to concrete game situations, grounding the LLM in consistent reference points across batches; (2) \emph{Constraint specification} enforces SPREAD, INDEPENDENCE, and COMPLETENESS as explicit requirements for feature discovery; (3) \emph{Batch formatting} uses numbered lists with fixed format (``\texttt{0: FEAT1=0.xx FEAT2=0.yy}''), enabling reliable parsing. The calibration anchors serve a dual role: they normalize scores across batches (analogous to few-shot examples) and provide interpretable semantics for each feature's scale.

\paragraph{Feature Families Across Games.}

Because feature discovery samples at temperature $0.7$, the LLM instantiates its features with different names on each run, but the underlying strategic axes are stable. Across seeds and both subgames, the recurring feature families are: current made-hand strength (e.g., \textsc{Current\_Made\_Strength}, the family most correlated with EHS at $r \approx 0.87$--$0.93$), nut potential (\textsc{Nut\_Ceiling} and variants), blocker effects (\textsc{Blocker\_Leverage}/\textsc{Blocker\_Denial}), river volatility and downside exposure, and domination / reverse-implied-odds risk. The made-strength family aligns with EHS, while the volatility, blocker, and domination families have substantially weaker EHS correlation (Table~\ref{tab:feature-ehs}), which is what lets the multi-dimensional abstraction separate hands that EHS would merge. After correlation filtering, 4--8 features survive per run (median 6).

\paragraph{Parse Rate and Robustness.}

Across all experiments (HUNL turn/flop/preflop, PLO4, and Riichi Mahjong), the overall parse rate exceeds 99\% (100\% on turn, preflop, PLO4, and Mahjong; 97.9\% on flop). The turn results are also robust to the exact discovery prompt: three prompt variants (rules-only, range-aware, and an explicit feature-discovery instruction) yield mean ratios that agree within the per-seed standard deviation at $K{\in}\{20,50\}$ (Appendix~\ref{app:prompt_robustness}), so the reported gains are not an artifact of a single prompt wording.

\section{Future Work}
\label{app:future_work}

Several directions follow directly from this work. First, characterizing the dependence on the LLM backbone: a frontier model is currently needed to beat EHS, and identifying which capabilities drive this gap would clarify when the pipeline transfers to weaker or specialized models. Second, a cheap proxy for abstraction quality that avoids solving the game would close the loop on oracle-free feature-set selection. Third, solver-in-the-loop refinement could use exploitability feedback to revise the discovered features online, and a principled fallback for unparsable scores would replace the current midpoint imputation.

More broadly, the pipeline instantiates a general recipe for rule-grounded LLM scoring: given a natural-language specification of evaluation criteria and calibration anchors, an LLM scores arbitrary objects on continuous dimensions, and downstream objective signals, here exploitability, verify the quality of those scores. This suggests two wider uses. First, LLM-as-a-judge: the recipe turns any domain whose criteria can be written down into a structured scorer without human preference data, and the verifiable-calibration view offers an alternative to agreement-with-human-labels as the quality signal for such judges. Second, LLM evaluation: the rules-to-features-to-clusters-to-exploitability chain is an automatic, zero-annotation benchmark of an LLM's ability to turn a specification into sound numerical knowledge, complementing game-playing benchmarks that probe decision quality (e.g., PokerBench \citep{zhuang2024pokerbench}) with a direct test of structured knowledge elicitation. Beyond scoring and evaluation, feature discovery is a step toward automated feature engineering, and the game-agnostic design transfers naturally to security games, auctions, and negotiation.